\documentstyle[epsfig,12pt]{article}
\newcommand{\be}{\begin{eqnarray}}
\newcommand{\ee}{\end{eqnarray}}

\date{ }
\begin{document}

\title{Baryon, Charged Hadron, Drell-Yan and $J/\psi$ 
Production in High Energy
Proton-Nucleus Collisions\thanks{Talks presented at the Workshop on
RHIC Physics and Beyond,  BNL, 23 Oct. 1998 (dedicated to
the memory of Klaus Kinder-Geiger). 
This paper is based on Phys. Rev. Lett. {\bf 82},
1626 (1999) and on preprint nucl-th/9812056.}
}

\author{Charles Gale$^{*}$\thanks{Second speaker.},
Sangyong Jeon$^{\dagger}$,
Joseph Kapusta$^{\#}$\thanks{First speaker.}\\ \\
\small \it 
 $^{*}$Physics Department,  McGill University\\
\small \it 
 Montreal, Quebec H3A 2T8, Canada\\
\small \it 
 $^{\dagger}$Nuclear Science Division,  Lawrence Berkeley National 
Laboratory\\ 
\small \it 
 Berkeley, CA 94720, USA\\ 
\small \it 
 $^{\#}$School of Physics and Astronomy,  University of Minnesota\\ 
\small \it 
 Minneapolis, MN 55455, USA}

\maketitle
\begin{abstract}

We show that the distributions of outgoing protons and charged hadrons in high 
energy
proton-nucleus collisions are described rather well by a linear extrapolation 
from proton-proton collisions.  The only adjustable parameter required is the 
shift in rapidity of a produced charged meson when it encounters a target 
nucleon.  Its fitted value is 0.16.  Next, we apply this linear extrapolation to 
precisely measured Drell-Yan cross sections for 800 GeV protons incident
on a variety of nuclear targets which exhibit a deviation from linear scaling in 
the atomic number $A$.  We show that this deviation can be accounted for by
energy degradation of the proton as it passes through the nucleus if
account is taken of the time delay of particle production due to
quantum coherence.  We infer an average proper coherence time of
0.4$\pm$0.1 fm/c, corresponding to a coherence path length of
8$\pm$2 fm in the rest frame of the nucleus. Finally, we apply the linear 
extrapolation to measured $J/\psi$ production cross 
sections for 200 and 450 GeV/c protons
incident on a variety of nuclear targets.  Our analysis takes into account 
energy loss of the beam proton, the time
delay of particle production due to quantum coherence, and absorption of the
$J/\psi$ on nucleons.  The best representation is obtained for a coherence
time of 0.5 fm/c, which is consistent with Drell-Yan production, and an 
absorption cross section of 3.6 mb, which is
consistent with the value deduced from photoproduction of
the $J/\psi$ on nuclear targets.

\end{abstract}

\clearpage

\section*{Introduction}

Propagation of a high energy particle through a medium is of interest
in many areas of physics.  High energy proton-nucleus scattering has
been studied for many decades by both the nuclear and particle physics
communities \cite{Wit}.  Such studies are particularly relevant for
the Relativistic Heavy Ion Collider (RHIC), which will collide beams
of gold nuclei at an energy of 100 GeV per nucleon, and for the Large
Hadron Collider (LHC), which will collide beams of lead nuclei at
1500 GeV per nucleon \cite{qm97}.

There are two extreme limits of a projectile scattering from a nucleus.
When the cross section of the projectile with a nucleon is very small,
as is the case for neutrinos, Glauber theory says that the cross section
with a nuclear target of atomic number $A$ grows linearly with $A$.  When
the cross section with an individual nucleon is very large, as is
the case for pions near the delta resonance peak, the nucleus appears
black and the cross section grows like $A^{2/3}$.  A more interesting
case is the production of lepton pairs with large invariant mass,
often referred to as Drell-Yan, in proton-nucleus
collisions.  Both the elastic and inelastic cross sections for
proton-nucleon scattering are relatively large, but the partial
cross section to produce a high mass lepton pair, being electromagnetic
in origin, is relatively small.  Experiments have shown that the
inclusive Drell-Yan cross section grows with $A$ to a power very
close to 1.  The theoretical interpretation is that the hard
particles, the high invariant mass lepton pairs, appear first
and the soft particles, the typical mesons, appear later
due to quantum-mechanical interference, essentially the uncertainty
principle.  These quantum coherence requirements also lead to the
Landau-Pomeranchuk-Migdal effect \cite{lpm}.  Deviations from the
power 1 by high precision Drell-Yan experiments \cite{E772}
at Fermi National Accelerator Laboratory (FNAL) suggest
that it may be possible to infer a finite numerical value for the
coherence time.  That is one of our goals.

Another of our goals is to take this coherence time effect into account when 
extracting an absorption cross section for $J/\psi$ on nucleons for $J/\psi$ 
particles produced in high energy proton-nucleus collisions.  This cross section 
is of great interest in the ongoing analysis and debate over whether quark-gluon 
plasma is formed in high energy nucleus-nucleus collisions.

Before addressing these goals it is imperative to have a basic description of 
high energy proton-nucleus collisions which reproduces the essential data on 
outgoing baryons and mesons.  We will use one particular theoretical approach, 
but it important to realize that any model or extrapolation which incorporates 
the same basic features will lead to the same conclusions we find here. 

\section*{Baryon and Charged Hadron Production}

For a basic description of high energy proton-nucleus scattering
we prefer to work with hadronic variables rather than partonic ones.
We make a straightforward linear extrapolation from proton-proton
scattering.  This extrapolation, referred to as LEXUS, was detailed
and applied to nucleus-nucleus collisions at beam energies of several
hundred GeV per nucleon in ref. \cite{lexus}.  Briefly, the
inclusive distribution in rapidity $y$ of the beam proton in an
elementary proton-nucleon collision is parameterized rather well by
\begin{equation}
W_1(y) = \lambda \frac{\cosh y}{\sinh y_0}
+ (1-\lambda)\delta(y_0-y) \, ,
\end{equation}
where $y_0$ is the beam rapidity in the lab frame.
The parameter $\lambda$ has the value 0.6 independent of beam
energy, at least in the range in which it has been measured, which
is $12-400$ GeV.  It may be interpretted as the fraction of all
collisions which are neither diffractive nor elastic.
As the proton cascades through the nucleus its
energy is degraded.  Its rapidity distribution satisfies an
evolution equation \cite{hwa84} whose solution is, after $i$ collisions
\cite{ck85}:
\begin{equation}
W_i(y) = \frac{\cosh y}{\sinh y_0} \sum_{k=1}^i
\left(
\begin{array}{c}
i\\k 
\end{array}
\right)
\frac{\lambda^k (1-\lambda)^{i-k}}{(k-1)!}
\left[ \ln\left( \frac{\sinh y_0}{\sinh y} \right) \right]^{k-1}
+ (1-\lambda)^i \delta(y_0-y) \, .
\end{equation}
This distribution then gets folded with impact parameter over the
density distribution of the target nucleus as measured by electron
scattering.

The net proton distributions are shown for 200 GeV protons incident on S and Au 
targets in Fig. 1.  The data is from NA35 \cite{NA35}.
\begin{figure}[htb]
\centerline{\epsfig{file=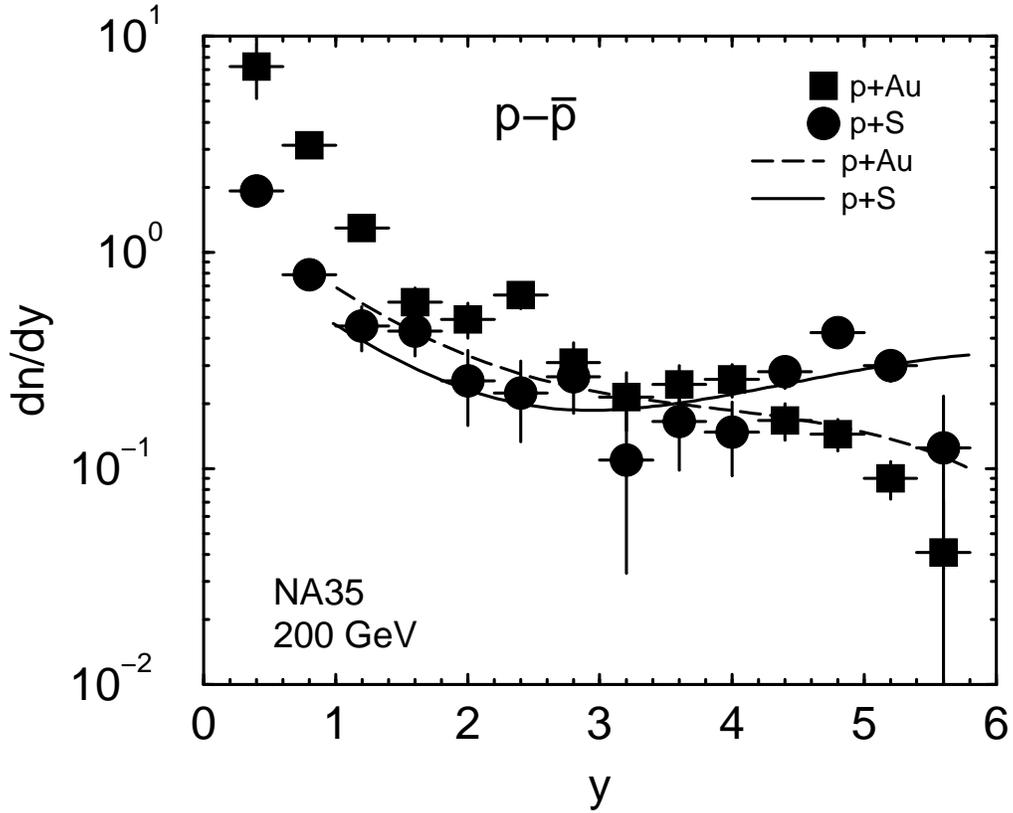,width=6.in}}
\vspace{-10pt}
\caption{The rapidity distribution of net protons in p+S and p+Au collisions at 
a beam energy of 200 GeV.  The data are from NA35 \protect\cite{NA35}. 
The curves are 
calculated with LEXUS with no free parameters.  The curves undershoot the data 
at low rapidity because target evaporation is not taken into account.}
\end{figure}
The theoretical curves 
fall below the data when the rapidity is less than 1, and for good reason.  The 
projectile deposits energy in the target which subsequently boils off low energy 
nucleons, but this effect is not included in the curves.  The minor discrepancy 
near the beam rapidity may be due to an inadequate parametrization of the 
elementary pp distribution eq. (1), or to experimental cuts and acceptances in 
momentum space, or to both.  Otherwise the agreement is very good and without 
free parameters.

Every inelastic collision produces an average number
of negatively charged hadrons given by the simple formula
\begin{equation}
\langle h^- \rangle_{NN} = 0.784 \,
 \frac{(\sqrt{s} - 2m_N -m_{\pi})^{3/4}}{s^{1/8}} \, .
\end{equation}
The negatively charged hadrons are Gaussian distributed in rapidity with
a width given by $\sqrt{\ln(\sqrt{s}/2m_N)}$ in customary notation.
The rapidity distribution of negatively charged hadrons, as computed
in the way described, is nearly centered in the nucleon-nucleon
center-of-momentum (c.m.) frame, whereas data taken for p+S and p+Au
collisions at 200 GeV \cite{NA35} are skewed towards the target rest frame.
If one allows for a small rapidity shift of 0.16 whenever a produced
hadron encounters a struck target nucleon, chosen with a sign corresponding
to a slowing down of the hadron relative to the nucleon, one obtains
the curves shown in Fig. 2.
\begin{figure}[htb]
\centerline{\epsfig{file=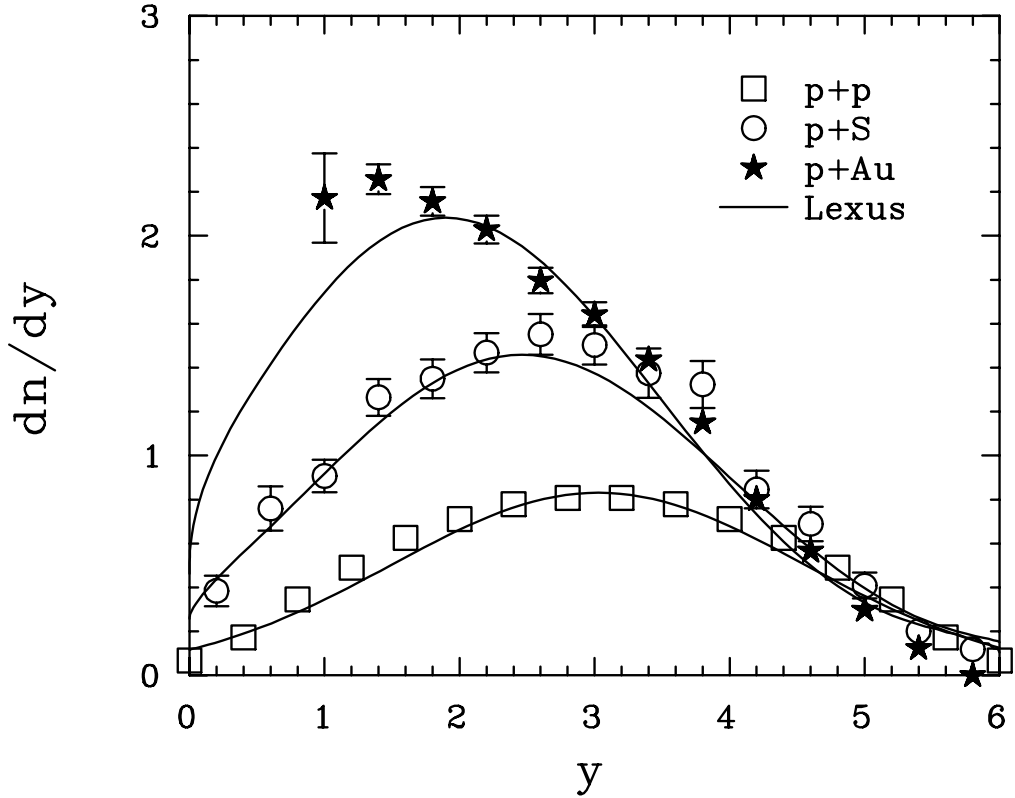,height=5.25in,angle=90}}
\vspace{15pt}
\caption{ The rapidity distribution of negatively charged hadrons
in pp, p+S, and p+Au collisions at a beam energy of 200 GeV.
The pp data are from \protect\cite{pp} and the p+S and p+Au data are
from NA35 \protect\cite{NA35}.  The curves are calculated with LEXUS with
a rapidity shift of 0.16 per struck nucleon. }
\end{figure}
The paucity of computed hadrons compared
to data at small rapidity in p+Au collisions is undoubtedly due to a
further cascading and particle production by struck nucleons in such a
large nucleus.  This physics could be incorporated by a more detailed
cascade code followed by nuclear evaporation, but is not essential
for our purposes in this paper.  Apart from that, the description
of the data is very good, including absolute normalization,
especially considering that there is only one free parameter.

\begin{figure}[h!]
\centerline{\epsfig{file=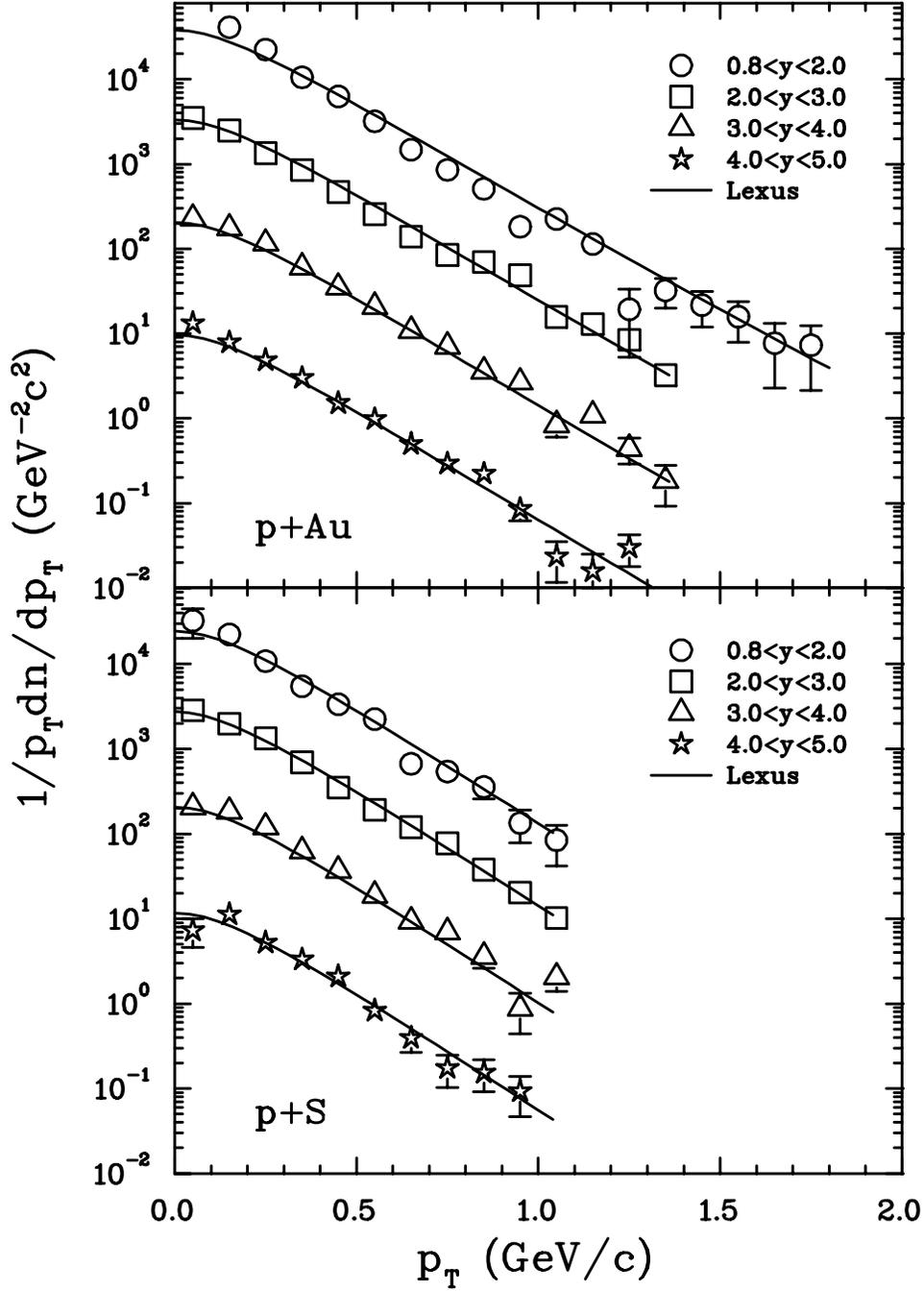,height=7.in}}
\vspace{15pt}
\caption{ The transverse momentum distributions for the same p+S
and p+Au collisions as in Fig. 2.  The curves are calculated with
LEXUS with a rapidity shift of 0.16 per struck nucleon. }
\end{figure}

As the proton cascades through the nucleus it undergoes a random walk
in transverse velocity.  This broadens the transverse momentum
distribution of the produced hadrons relative to pp collisions in the
way described in ref. \cite{lexus}.  The transverse momentum
distributions, for various windows of rapidity, are shown in
Fig. 3.  There are no free parameters apart from the rapidity
shift which was already fitted to be 0.16.

\section*{Drell-Yan Production}

Having satisfied ourselves that we have a reasonable quantitative
description of the soft hadronic physics, we now turn to a
description of the Drell-Yan.  Figure 4 is a schematic
of two limits and an intermediate situation.  
\begin{figure}[h!]
\centerline{\epsfig{file=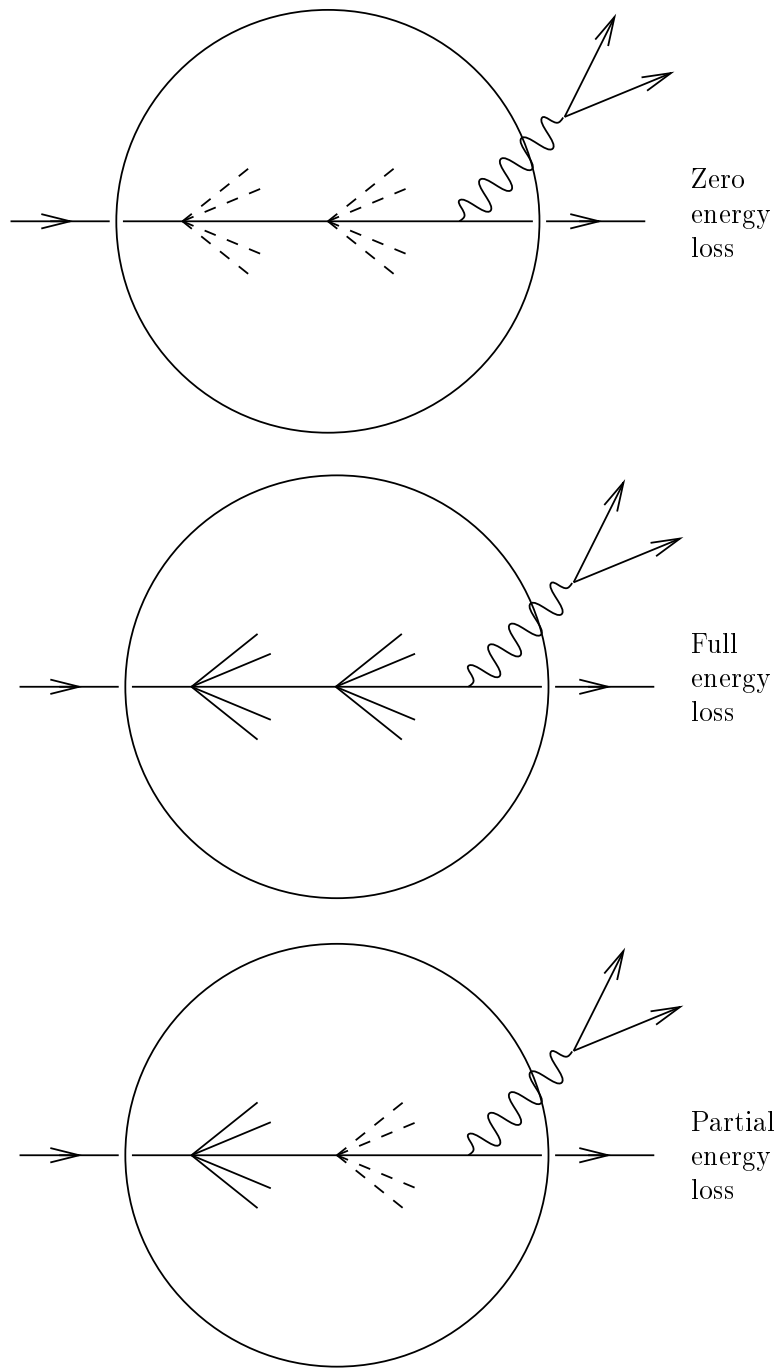,width=4.in}}
\vspace{15pt}
\caption{ Schematic of a high energy proton passing through a nucleus.
The upper panel represents full energy loss: hadrons are produced
immediately and the proton has less energy available for each subsequent
collision.  The middle panel represents partial energy loss: there is
a finite time delay before hadrons are produced and so the proton
has more energy available to create a high energy Drell-Yan pair.
The bottom panel represents no energy loss: the usual Glauber picture. }
\end{figure}
One limit is full
energy degradation of the proton as it traverses the nucleus.  Produced
hadrons appear immediately with zero coherence time, causing the proton
to have less energy available to produce a Drell-Yan pair at the backside
of the nucleus.  The other limit is usually referred to as Glauber,
although this is a bit of a misnomer.  Produced hadrons, being soft
on the average, do not appear until after the hardest particles, the
Drell-Yan pair, have already appeared.  This is the limit of a very
large coherence time, and it allows the proton to produce the Drell-Yan
pair anywhere along its path with the full incident beam energy.
An intermediate case is one of finite, nonzero coherence time.
By the time the proton wants to make a Drell-Yan pair on the backside
of the nucleus, hadrons have already appeared from the first collision
but not from the second.  Therefore the proton has more energy available
%\clearpage
\noindent to produce the Drell-Yan pair than full zero coherence time but less
energy than with infinite coherence time.  This ought to result
in an $A$ dependence less than 1, with the numerical value determined
by the coherence time.  It is instructive to contemplate the relative
importance of energy loss and coherence time for an 800 GeV proton
incident on a {\em very} large nucleus, such as a neutron star: Can
one imagine the proton reaching the backside of a neutron star and
producing a Drell-Yan pair without having suffered {\em any} energy loss?

Consistent with our philosophy to describe everything in terms of
hadronic variables we should use a parametrization of measured
Drell-Yan cross sections in pp and pn collisions.  However, we
need these over a very broad energy range because of the decreasing
energy of the proton as it cascades through the nucleus, and such
broad measurements have not been made.  Therefore, we compute the
Drell-Yan yields in individual pp and pn collisions using the parton
model with the GRV structure functions \cite{grv94} to leading order
with a K factor.  These structure functions distinguish between pp
and pn collisions.  We have compared the results to pp collisions at
the same beam energy of 800 GeV \cite{800pp} and
found the agreement to be excellent for all values of $x_F$.

The experiment E772 \cite{E772} measured the ratio
$\sigma^{\rm DY}_{pA}/(\sigma^{\rm DY}_{pd}/2)$.  Were there no energy
loss and all nuclei were charge symmetric this ratio would be
equal to $A$.  The experiment measured muon pairs with invariant mass
$M$ between 4 and 9 GeV and greater than 11 GeV to eliminate
the $J/\psi$ and $\Upsilon$ contributions.  The data has been presented
in 7 bins of Feynman $x_F$ from 0.05 to 0.65.  (Recall that $x_F$ is
the ratio of the muon pair longitudinal momentum to the incident beam
momentum in the nucleon-nucleon c.m. frame.)  Data for exemplary values
of $x_F$ are shown in Figs. 5 to 7.
\begin{figure}[ht]
\centerline{\epsfig{file=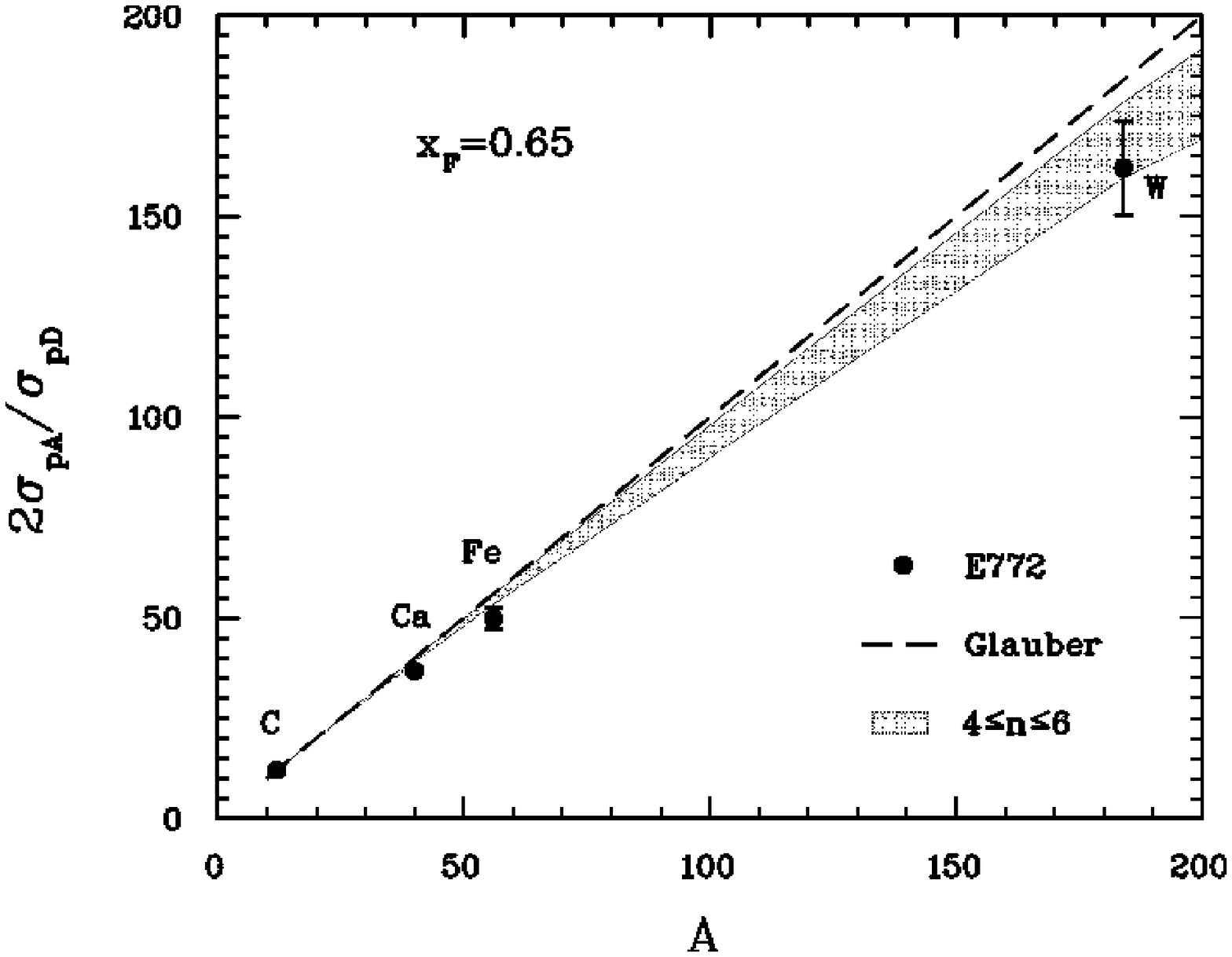,height=4.5in,angle=90}}
\vspace{15pt}
\caption{ The ratio of the pA Drell-Yan cross section to the
proton-deuterium cross section divided by 2 for a beam energy
of 800 GeV.  The data are from E772 \protect\cite{E772}.  The dashed
line assumes a scaling linear in the atomic number A.  The shaded
region represents our calculations with a coherence time ranging
from 4 to 6 proton-nucleon collisions, both elastic and inelastic.
We computed with target nuclei C, Ca, Fe, W and Pb and interpolate
between with straight lines to guide the eye.  The value of Feynman
$x_F$ of the Drell-Yan pair is 0.65. }
\end{figure}
The data should fall on
the dashed line if the ratio of cross sections is $A$.  There is
a small but noticeable departure for tungsten and at the largest value
of $x_F$.  This is to be expected if energy loss plays a role as it
must affect the largest target nucleus and the highest energy muon
pairs the most \cite{enhance}.

We have computed the individual cross sections $\sigma^{\rm DY}_{pA}$
with a variable time delay.  The proton cascades through the nucleus
as described earlier, but we assume that the energy available to
produce a Drell-Yan pair is that which the proton has after $n$
previous collisions.  Thus $n = 0$ is full energy loss and $n = \infty$
is zero energy loss.  We have taken the resulting proton-nucleus
Drell-Yan cross section, multiplied it by 2, divided it by the
sum of the computed pp and pn cross sections and display the results
in Figs. 5 to 7.  The lower edge of the shaded regions in the figure
corresponds to $n = 4$ and the upper edge to $n = 6$.
Overall the best representation of the data lies in this
range.  This collision number shift is easily converted to a coherence
time.  Let $\tau_0$ be the coherence time in the c.m. frame of the
colliding nucleons.  This is essentially the same as the formation
time of a pion since most pions are produced with rapidities near
zero in that frame.  The first proton-nucleon collision is the most
important, so boosting this time into the rest frame of the target
nucleus and converting it to a path length (proton moves essentially
at the speed of light) gives $\gamma_{\rm cm}\,c\,\tau_0 \approx
\sqrt{\gamma_{\rm lab}/2}\,c\,\tau_0$.  This path length may then be
equated with $n$ times the mean free path
$l = 1/\sigma^{\rm tot}_{\rm NN} \rho$.  Using a total cross section
of 40 mb and a nuclear matter density of 0.155 nucleons/fm$^3$ we
obtain a path length of 8$\pm$2 fm and a proper coherence time
of 0.4$\pm$0.1 fm/c corresponding to $n=5\pm 1$.

This value of the proper coherence time is just about what should
have been expected {\it a priori}.  In the c.m. frame of the colliding
nucleons at the energies of interest a typical pion is produced
with an energy of E$_{\pi} \approx 500$ MeV.  By the uncertainty
principle this takes a time of order $\hbar c/E_{\pi} \approx 0.4$
fm/c.

%\clearpage
\begin{figure}[h!]
\centerline{\epsfig{file=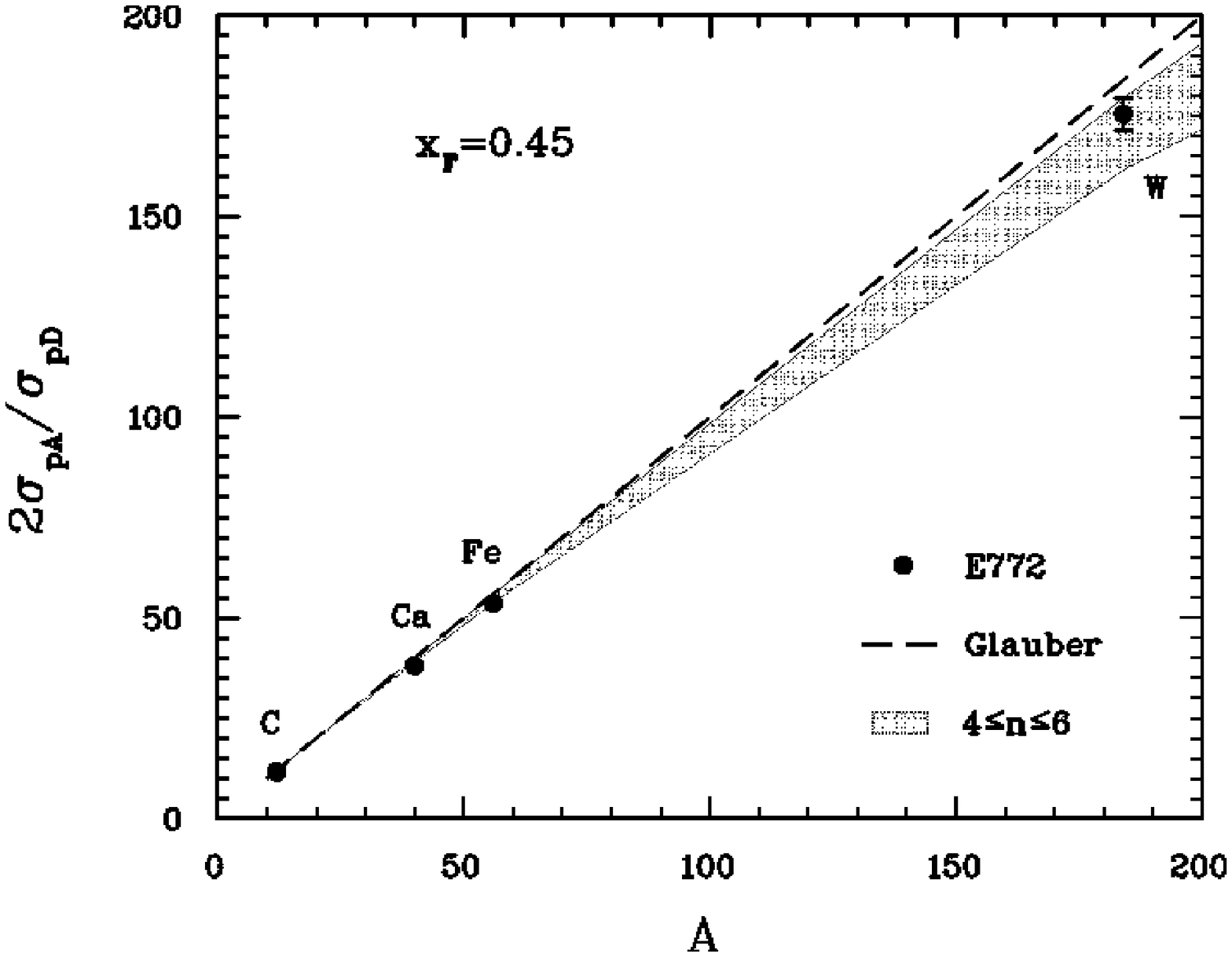,height=4.5in,angle=90}}
\vspace{10pt}
\caption{ Same as Fig. 5 but for $x_F = 0.45$.}
\vspace{0.25in}
\centerline{\epsfig{file=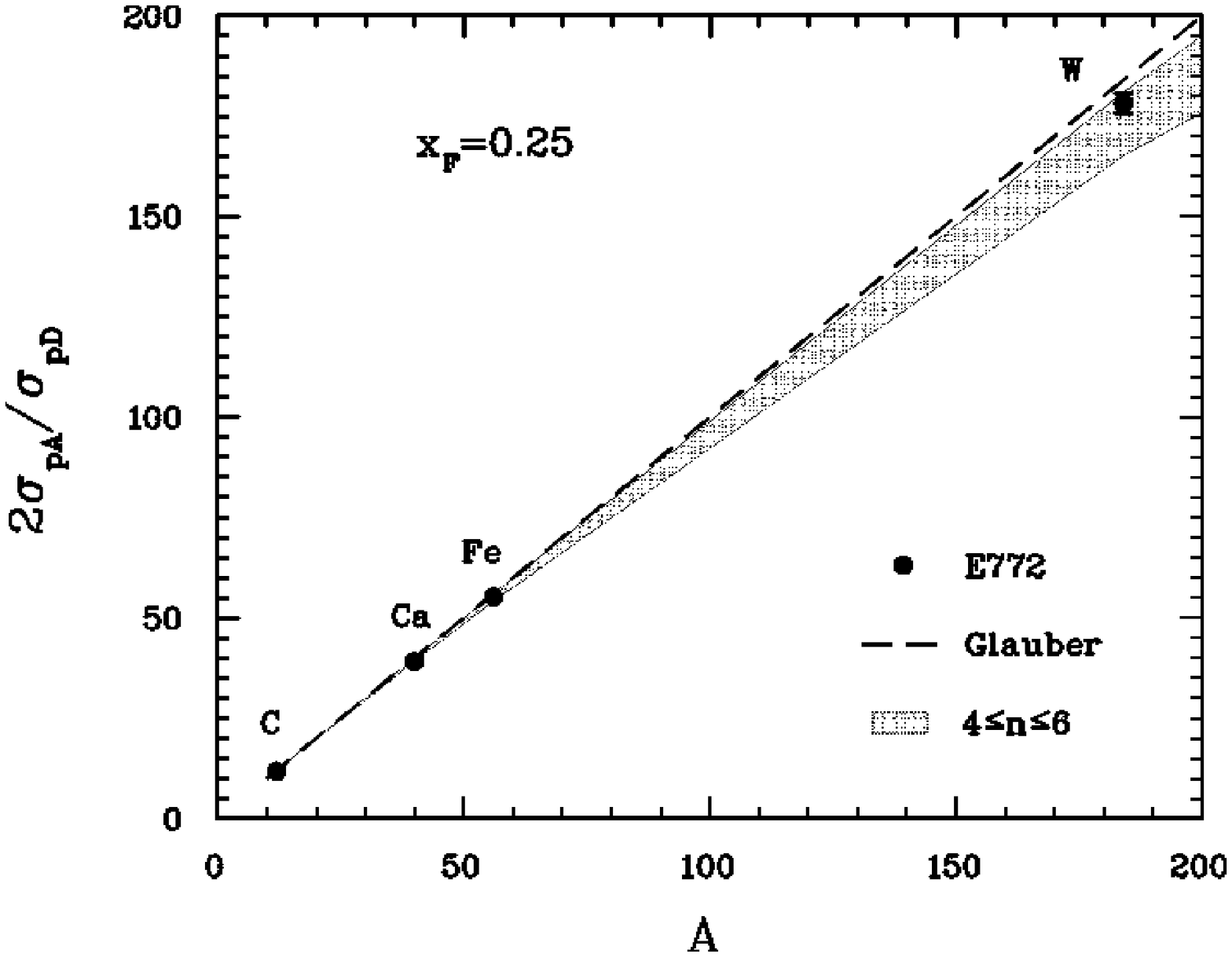,height=4.5in,angle=90}}
\vspace{10pt}
\caption{ Same as Fig. 5 but for $x_F = 0.25$.}
\end{figure}
%\clearpage

\section*{$J/\psi$ Production and Absorption}

A process related to Drell-Yan is the production of $J/\psi$ which we shall 
address now.  This is also a relatively
hard process and so both energy loss of the beam proton and the
Landau-Pomeranchuk-Migdal effect must be taken into account. However, there
is an additional effect
which plays a role, and that is the occasional absorption or breakup of the
$J/\psi$ in encounters with target nucleons.  (The inelastic interaction of one
of the leptons in Drell-Yan production with target nucleons is ignorably
small.)  The absorption cross section, $\sigma_{\rm abs}$, has been estimated
in a straightforward Glauber analysis without energy loss and with an infinite
coherence/formation time to be about 6-7 mb \cite{DimaJ}.  This has formed the
basis for many analyses of $J/\psi$ suppression in heavy ion collisions.  Any
anomalous suppression may be an indication of the formation of quark-gluon
plasma \cite{qm97,matsui}, hence the importance of obtaining the most accurate
value of $\sigma_{\rm abs}$ possible.  This cross section has also been
inferred from photoproduction experiments of $J/\psi$ on nuclei from which a
value much less than that has been obtained \cite{photo}.  This has been a
puzzle. One attempt to resolve this apparent discrepancy consists of
modeling the produced $J/\Psi$ state as a pre-resonant color dipole state
with two octet charges \cite{dima2}; however, the results are only
semi-quantitative.
\begin{figure}[h]
\centerline{\epsfig{file=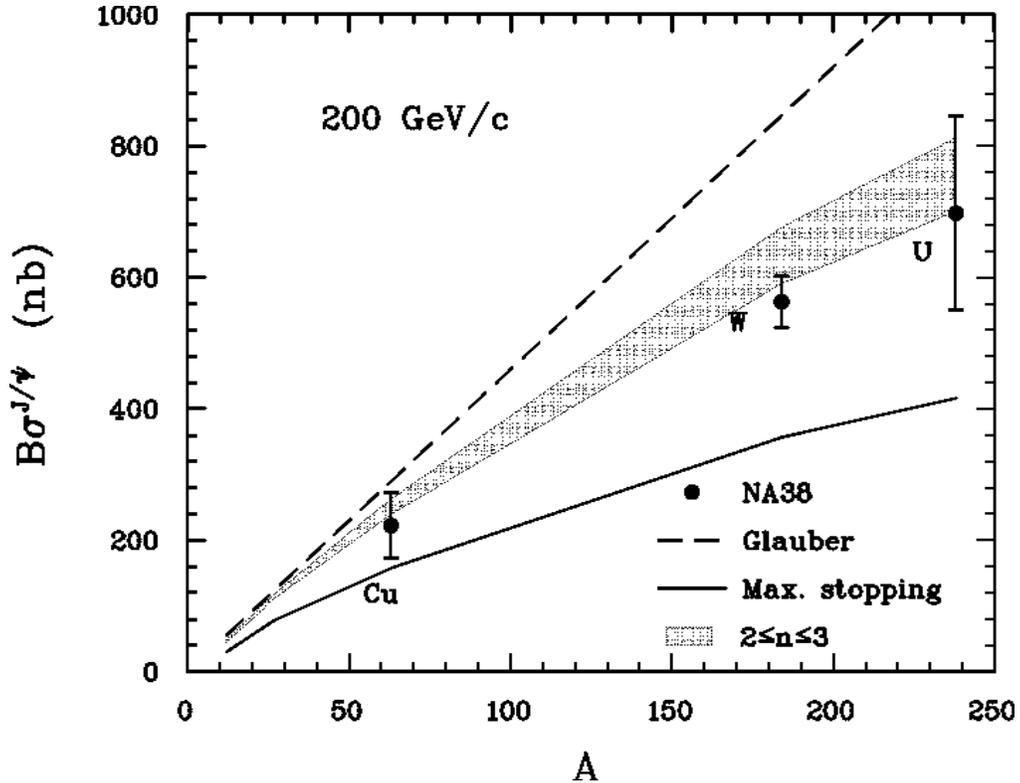,height=5.25in,angle=90}}
\vspace{15pt}
\caption{ Branching ratio into muons times cross section to produce $J/\psi$
with $x_F \ge 0$ in proton-nucleus collisions at 200 GeV/c.  The data is from
NA38 \protect\cite{Carlos,NA38}.  The dashed line is $A$ times the
nucleon-nucleon
production cross section.  The solid curve represents full energy
loss with zero coherence/formation time, while the banded region represents
partial energy loss
with a coherence/formation time within the limits set by Drell-Yan production.
(Computations were done for C, Al, Cu, W and U and the points connected by
straight lines to guide the eye.)}
\end{figure}

In order to compute the
production cross section of $J/\psi$ in proton-nucleus collisions we need a
parametrization of it in the more elementary nucleon-nucleon collisions. For
this we call upon the parametrization of a compilation of data by
Louren\c{c}o \cite{Carlos}.
\begin{equation}
B \sigma_{NN \rightarrow J/\psi}(x_F > 0) = 37 \left( 1 - m_{J/\psi}/\sqrt{s}
\right)^{12} \, {\rm nb}
\end{equation}
Here $B$ is the branching ratio into dimuons and $x_F$ is the ratio of the
momentum carried by $J/\psi$ to the beam momentum in the center of mass frame
$(-1 < x_F < 1)$.  Due to the degradation in momentum of the proton as it
traverses the nucleus it is important to know the $x_F$ dependence of the
production.  The Fermilab experiment E789 has measured this dependence at 800
GeV/c \cite{E789} to be proportional to $(1-|x_F|)^5$.  Assuming that this
holds at lower energy too we use the joint $\sqrt{s}$ and $x_F$
functional dependence and magnitude:
\begin{equation}
\frac{d\sigma_{NN \rightarrow J/\psi}}{dx_F} = 6 \sigma_{NN \rightarrow
J/\psi}(x_F > 0) (1-|x_F|)^5 \, .
\end{equation}
The cross section in proton-nucleus collisions can now be computed in LEXUS
with no ambiguity.

Figures 8 and 9 show the results of our calculation in comparison to data taken
by NA38 \cite{NA38} and NA51 \cite{NA51}, respectively.
\begin{figure}[h]
\centerline{\epsfig{file=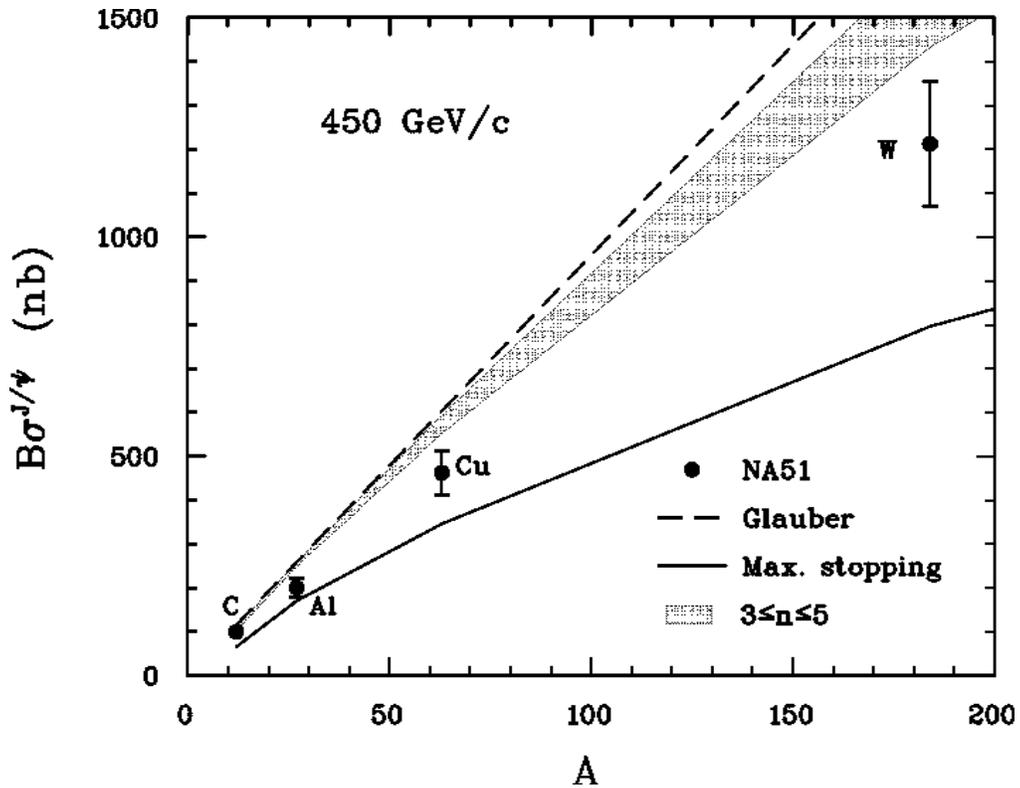,height=5.25in,angle=90}}
\vspace{15pt}
\caption{ Same as figure 8 but for a beam momentum of 450 GeV/c.  Data is from
NA51 \protect\cite{Carlos,NA51}. }
\end{figure}
The dashed curves
are $A$ times the nucleon-nucleon production cross section; they obviously
overestimate the data.  The
solid curves show the result of LEXUS with full energy degradation of the beam
proton without account taken of the Landau-Pomeranchuk-Migdal effect; they
obviously underestimate the data.  The hatched regions represent the inclusion
of the latter effect with a proper formation/coherence time $\tau_0$ in the
range of 0.3 to 0.5 fm/c consistent with Drell-Yan production.  The
time delay is implemented as described previously:  The energy available for the
production of $J/\psi$ is that which the proton had $n$ collisions prior; that
is, the previous $n$ collisions are ignored for the purpose of determining the
proton's energy. This is an approximate treatment of the
Landau-Pomeranchuk-Migdal effect.  The $n$ is related to the beam energy and
to  the coherence time $\tau_0$ in the center of mass frame of the
colliding nucleons.  Using, as before, a total cross section
of 40 mb, a nuclear matter density of 0.155 nucleons/fm$^3$, and
$0.3 < \tau_0 <
0.5$ fm/c we obtain $2 < n < 3$ at 200 GeV/c and $3 < n < 5$ at 450 GeV/c. As
may be seen from the figures, the data is overestimated, indicating the
necessity for nuclear absorption.

We now introduce a $J/\psi$ absorption cross section on nucleons and compute
its effect within LEXUS in the canonical way \cite{DimaJ}.  When the $J/\psi$
is created there will in general be a nonzero number of nucleons blocking its
exit from the nucleus.  Knowing where the $J/\psi$ is created allows one to
calculate how many nucleons lie in its path, and hence, to compute the
probability that it will be dissociated into open charm.  We choose a value
of $\tau_0$ allowed by Drell-Yan measurements, mentioned above, and then
vary $\sigma_{\rm abs}$, assuming that it is energy
independent.  The lowest value of chi-squared for the 200 and 450 GeV/c
data set taken together is obtained with $\tau_0 = 0.5$ fm/c and
$\sigma_{\rm abs} = 3.6$ mb.  The results are shown in figures 10 and 11.
\begin{figure}[ht!]
\centerline{\epsfig{file=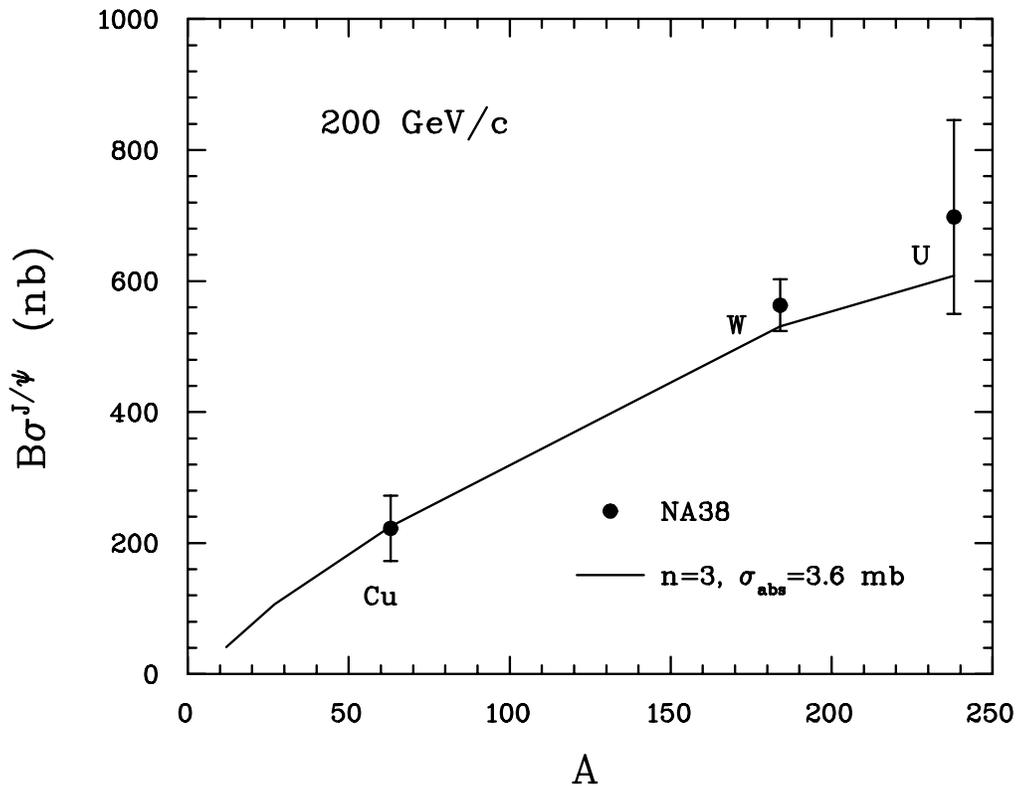,height=5.25in,angle=90}}
\vspace{15pt}
\caption{ Same data as in figure 8.  The solid curve is the best fit of the
model which includes beam energy loss with a coherence time of 0.5 fm/c (n=3 at
this energy) and a $J/\psi$ absorption cross section of 3.6 mb. }
\end{figure}
\begin{figure}[h!]
\centerline{\epsfig{file=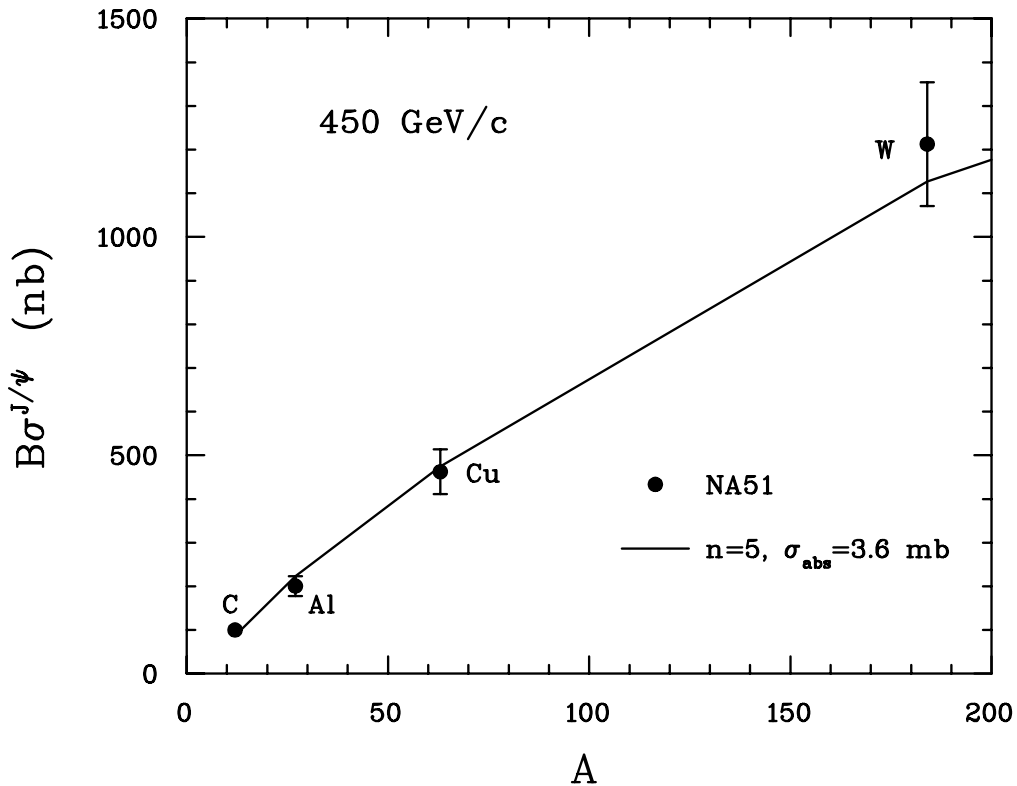,height=5.25in,angle=90}}
\vspace{15pt}
\caption{ Same data as in figure 9.  The solid curve is the best fit of the
model which includes beam energy loss with a coherence time of 0.5 fm/c (n=5 at
this energy) and a $J/\psi$ absorption cross section of 3.6 mb. }
\end{figure}
The fitted values all lie within one standard deviation of the data
points.  This is quite a satisfactory representation of the data.  It means
that both Drell-Yan and $J/\psi$ production in high energy proton-nucleus
collisions can be understood in terms of a conventional hadronic
analysis when account is taken of the energy loss of the beam proton, the
Landau-Pomeranchuk-Migdal effect, and nuclear absorption of the $J/\psi$
in the final state.  It also means that the absorption cross section for
$J/\psi$ inferred from high energy proton-nucleus collisions is
consistent with the value inferred from photoproduction experiments on nuclei.

\section*{Conclusion}

The analysis performed here can and should be improved upon.  What
we have done is a rough approximation to adding the quantum mechanical
amplitudes for a proton scattering from individual nucleons within
a nucleus.  A more sophisticated treatment would undoubtedly lead
to even better agreement with experiment, but the inferred value of
the proper coherence time is unlikely to be much different than obtained
with this first estimate.  It will be very instructive to repeat this
analysis in the language of partonic variables.  Actually, the analysis
with parton energy loss alone was reported by Gavin and Milana \cite{sean}
with satisfactory results obtained for Drell-Yan if the quarks/antiquarks
lose about 1.5 GeV/fm.  Nuclear shadowing \cite{shadows} needs to be taken into 
account too.  The
relationship among all these effects is not well-understood, nor is the
relationship between these effects in partonic and hadronic variables.
Finally, the implications for nucleus-nucleus collisions
\cite{kaha2} will undoubtedly be important; they are under investigation.

\section*{Acknowledgements}

This work was supported by the U. S. Department of Energy under grants
DE-FG02-87ER40328, DE-AC03-76SF00098 and DE-FG03-93ER40792, by the Natural 
Sciences and Engineering Research Council of Canada, and by the Fonds FCAR of 
the Quebec Government.


\begin{thebibliography} {20}
\bibitem{enhance} There is actually an enhancement of this ratio
near $x_F = 0$, and possibly for negative $x_F$ although it hasn't
been measured by E772 in the backward direction.  Such an
enhancement may be expected due to Drell-Yan production in occasional
collisions of secondary particles with target nucleons.
\end{thebibliography}
\end{document}